\documentclass[9pt,twocolumn]{revtex4-1}
\usepackage{amsmath}
\usepackage{amssymb}
\usepackage{graphicx}
\usepackage{bm}
\usepackage{lineno}
\usepackage{xcolor,soul}
\usepackage{subfigure}
\usepackage{epstopdf}
\usepackage{amsthm}
\usepackage{color}
\usepackage{soul}

\makeatletter
\renewcommand{\section}{\@startsection{section}{1}{0mm}
  {-\baselineskip}{0.5\baselineskip}{\bf\leftline}}
\makeatother
\makeatletter
\renewcommand{\subsection}{\@startsection{subsection}{1}{0mm}
  {-\baselineskip}{0.5\baselineskip}{\bf\leftline}}
\makeatother

\begin{document}

\title{Optimal phase measurements in a lossy Mach-Zehnder interferometer}
\author{Wenfeng Huang,$^{1}$ Xinyun Liang,$^{1}$ Chun-Hua Yuan,$^{1}$ Weiping Zhang,$^{2,3,4}$ and L.Q.Chen$^{1,*}$}
\address{$^{1}$State Key Laboratory of Precision Spectroscopy, Quantum
Institute for Light and Atoms, Department of Physics, East China Normal
University, Shanghai 200062, China\\
$^{2}$School of Physics and Astronomy, Tsung-Dao Lee Institute, Shanghai Jiao Tong University,
Shanghai 200240, People’s Republic of China\\
$^{3}$Collaborative Innovation Center of Extreme Optics, Shanxi University, Shanxi 030006, People’s Republic of China\\
$^{4}$Shanghai Research Center for Quantum Sciences, Shanghai 201315, People’s Republic of China}
\maketitle

\section*{abstract}
In this work, we discuss two phase-measurement methods for the Mach-Zehnder interferometer (MZI) in the presence of internal losses and give the corresponding optimum conditions. We find theoretically that when the core parameters (reflectivities, phase difference) are optimized, the phase sensitivity of the two methods can reach a generalized bound on precision: standard interferometric limit (SIL). In the experiment, we design an MZI with adjustable beam splitting ratios and losses to verify phase sensitivity optimization. The sensitivity improvements at loss rates from 0.4 to 0.998 are demonstrated based on difference-intensity detection, matching the theoretical results well. With a loss up to 0.998 in one arm, we achieve a sensitivity improvement of 2.5 dB by optimizing reflectivity, which equates to a 5.5 dB sensitivity improvement in single-intensity detection. Such optimal phase measurement methods provide practical solutions for the correct use of resources in lossy interferometry.

\section*{Introduction}

Mach-Zehnder Interferometer (MZI), one of the frequently-used optical
interferometers \cite{23}, is capable of ultra-sensitive phase measurements of various physical quantities, such as length \cite{24}, time \cite{25},
etc. And it has been applied in many regions including optical
telecommunications \cite{4}, quantum information \cite{5}, quantum
entanglement \cite{6}, quantum logic \cite{7}, etc. The phase sensitivity
\cite{1,2,3} 
is the critical parameter in the practical application of MZI. 

However, the losses in the light path, especially the unbalanced losses in two
interference arms \cite{12}, always bring the vacuum noise \cite{11} and
reduce the sensitivity in the phase measurements\cite{8,9,10}. Unbalanced losses exist commonly in nonlinear
processes \cite{17,18}, MZI \cite{19,20}, fiber optical communication \cite%
{21}, and especially gravitational wave measurements in space \cite{26}. For
instance, unbalanced losses are unavoidable for multipass interferometry, whose one interference arm passes
through the sample multiple times to obtain a multiple of the phase shift \cite{13,14,15,16}. In addition, in
the LISA proposal \cite{27,28}, the signal interference arm suffering
significant propagation loss is designed to return and interfere with
the lossless local reference arm. How to improve the phase sensitivity with
the existence of substantial unbalanced losses is crucial for the practical
application of MZI. \qquad

The optimization of phase estimation with internal losses has been discussed in Ref.\cite{29,30}. The quantum Fisher information (QFI) and its associated
quantum Cramér-Rao bound (QCRB) reveal the benchmark for precision in lossy interferometry. According to the QCRB, the first beam split ratio should be unbalanced to improve the precision when internal losses are unbalanced, regardless of the detection method used.
This generalized bound on precision is defined as standard interferometric limit (SIL). However, they do not provide specific methods for optimization in practical applications.

In this paper, we demonstrate optimization of the phase sensitivity in MZI with the
existence of unbalanced losses by adjusting the beam splitting ratio. The
optimal conditions are given for respective single-intensity and
difference-intensity detections. In theory, the optimal phase sensitivity of both
detections can reach the ultimate limit of such a lossy interferometer, that
is SIL, by properly allocating resources in two arms. With the same input states, the improvement of the phase sensitivity increases with the loss
rate. Further, we experimentally design the MZI with variable beam splitting ratios using polarization beam splitters and half-wave plates
and demonstrate the sensitivity improvement at loss rates from 0.4 to 0.998. 
Compared to conventional MZI (CMZI, constructed by two balanced 50:50 beam splitters), a phase sensitivity improvement of 2.5 dB is achieved using different-intensity detection with optimal beam splitting ratio $R_{1}^{opt}=0.04$ when the loss is 0.998. This result is equivalent to achieving a phase sensitivity improvement of 5.5 dB in intensity detection.
This MZI optimization scheme should be useful for applications with significant unbalanced loss.
\begin{figure}%"[]"中为位置参数，四个参数tbph依次是置顶、置底、浮动、当前位置，，选用的参数优先顺序为h-t-b-p
	\centering
	\includegraphics[width=1\columnwidth]{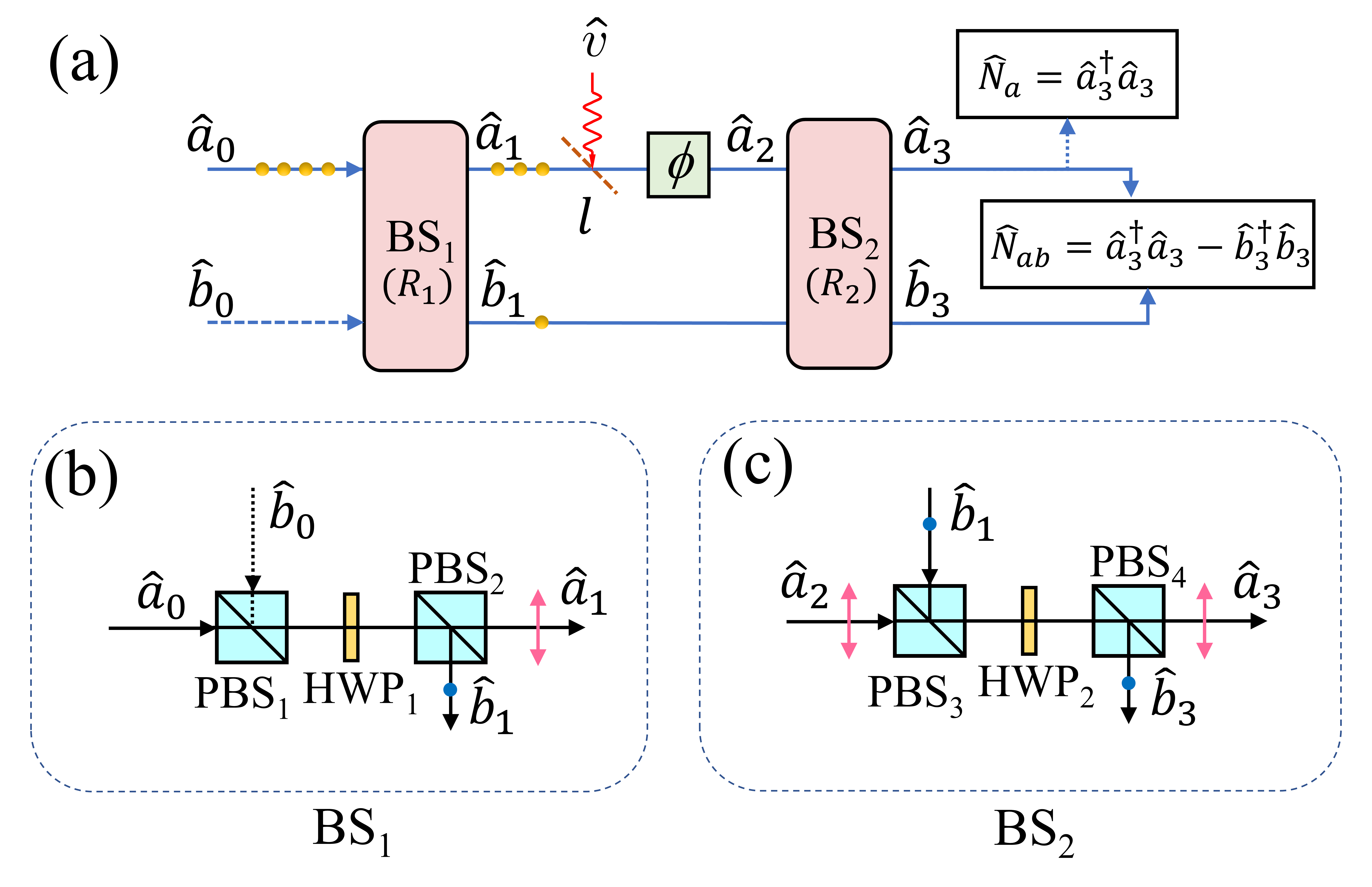}
	\caption{\textbf{(a)} Lossy MZI model with single-intensity detection and difference-intensity detection. The arm $a$
		experiences the loss (loss rate: $l$) and phase shift $\protect\phi $. 
		$\hat{a}_{0}$: annihilation operators of coherent state; $\hat{b}_{0},\hat{v}$: annihilation operators of vacuum state. $\hat{a}_{i},\hat{b}_{i}$ $(i=1,2,3) $: annihilation operators for beams at different positions of the interferometer. The yellow balls show the distribution of photons. \textbf{(b-c)} Schematic of beam splitters.BS$_{1,2}$: beam splitters with respective
		reflectivities $R_{1,2}$. PBS: polarization beam splitter; HWP: half-wave
		plate.}
	\label{fig.1}
\end{figure}

\section*{Theory}

In Fig. \ref{fig.1}(a), We consider an MZI model composed of the two input fields (a coherent field and a vacuum field) and
two beam splitters BS$_{1,2}$ with reflectivities $R_{1,2}$. An adjustable
attenuator with a loss rate of $l$ is placed in a single arm $a$ to simulate the
propagation loss (the case of losses on two arms has been discussed in Appendix \ref{B}). The input-output relation of BS$_{1}$ is given as
\begin{equation}
	\begin{aligned}
		\hat{a}_{1} =\sqrt{1-R_{1}}\hat{a}_{0}+i\sqrt{R_{1}}\hat{b}_{0}, \\
		\hat{b}_{1} =\sqrt{1-R_{1}}\hat{b}_{0}+i\sqrt{R_{1}}\hat{a}_{0},
	\end{aligned}
\end{equation}%
where $\hat{a}_{0}$ and $\hat{b}_{0}$ are the annihilation operators of the coherent and vacuum fields, respectively.
We treat the loss as a fictitious beam splitter, $\hat{v}$ is the annihilation operator of the vacuum state introduced by the
loss. After experiencing internal loss $l$ and considering a phase shift $\phi $
in the lossy arm $a$, $\hat{a}_{1}$ becomes
\begin{equation}
	\hat{a}_{2}=(\sqrt{1-l}\hat{a}_{1}+i\sqrt{l}\hat{v})e^{i\phi }.
\end{equation}%
The phase shift can be obtained from the interference outputs $\hat{a}_{3},\hat{b}_{3}$ by single-intensity detection or
difference-intensity detection \cite{22}. Below, we theoretically analyze
the phase sensitivity of MZI with these two detections.
\begin{figure*}
	\centering
	\includegraphics[width=2\columnwidth]{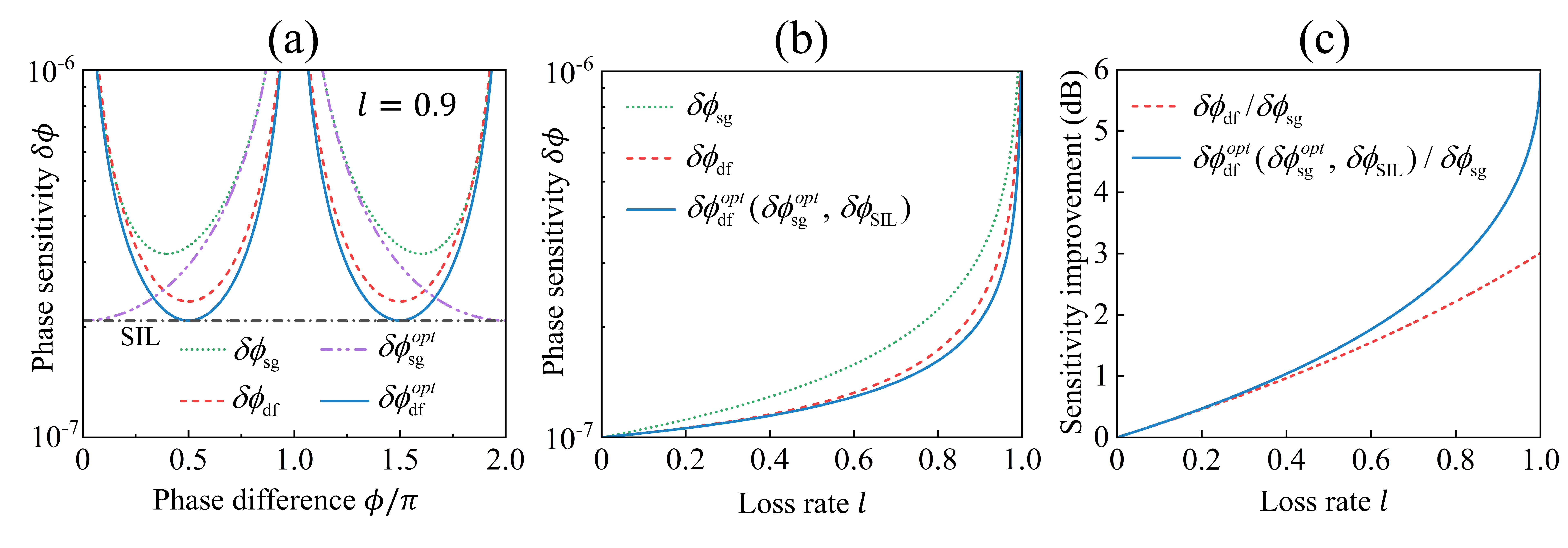}
	\caption{\textbf{Theoretical optimization analysis of $\text{sg}$- and $\text{df}$-detections.} \textbf{(a)} Phase sensitivity of two detections versus phase difference $\phi$ with balanced beam splitting ratios ($\delta\phi _{\text{sg}}$, $\delta\phi _{\text{df}}$) or optimal ones ($\delta\phi _{\text{sg}}^{opt}$, $\delta\phi _{\text{df}}^{opt}$) in MZI. Loss rate $l=0.9$; Photon number $N=10^{14}$. \textbf{(b)} Phase sensitivity $\protect\delta \protect\phi $ as a function of the loss rate $l$. All results are obtained with the respective best phase difference.
		$\delta\phi_{\text{SIL}}$: sensitivity of standard interferometric limit. \textbf{(c)} Sensitivity improvements are defined as $-20\text{log}_{10}(\delta \phi/\delta\phi _{\text{sg}})$, with the non-optimized sensitivity of $\text{sg}$-detection $\delta\phi _{\text{sg}}$ as the reference. $\protect\delta \protect\phi$ represents $\delta\phi _{\text{df}}$, $\delta\phi _{\text{sg}}^{opt}$, $\delta\phi _{\text{df}}^{opt}$, or $\delta\phi_{\text{SIL}}$. The $\phi _{\text{sg}}^{opt}$, $\delta\phi _{\text{df}}^{opt}$, and  $\delta\phi_{\text{SIL}}$ curves overlap together in (b) and (c).}
	\label{fig.2}
\end{figure*}
\bigskip

\textbf{I. Single-intensity (\text{sg}) detection}

Single-intensity detection only measures $\hat{N}_{a}=\hat{a}_{3}^{\dag }%
\hat{a}_{3}$ of the output $a_{3}$ using one detector. The phase sensitivity
$\delta \phi _{\text{sg}}$ is evaluated as [details see Appendix \ref{B1}.1]

\begin{equation}
	\begin{aligned}
		\delta \phi _{\text{sg}} =\frac{\left\langle \Delta \hat{N}_{a}\right\rangle }{\left\vert \partial \left\langle \hat{N}_{a}\right\rangle /\partial \phi\right\vert }=\dfrac{\sqrt{(B-C\cos \phi )N}}{CN\left\vert \sin\phi \right\vert},
	\end{aligned}
\end{equation}%
with
\begin{equation}
	\begin{aligned}
		&B =(1-R_{1})(1-R_{2})(1-l)+R_{1}R_{2},& \\
		&C =2\sqrt{(1-R_{1})(1-R_{2})R_{1}R_{2}(1-l)},&
	\end{aligned}
\end{equation}%
where $N$ is the photon number of the coherent optical field $a_{0}$. The
optimal phase shift $\phi ^{opt}$ and optimal reflectivities $%
R_{1}^{opt},R_{2}^{opt}$ corresponding to the best phase sensitivity are
\begin{equation}
	\begin{aligned}
		&R_{1}^{opt}=R_{2}^{opt}=\frac{\sqrt{1-l}}{1+\sqrt{1-l}},&\\
		&\phi^{opt}=0,2\pi.&
	\end{aligned}
\end{equation}%
With $R_{1}^{opt},R_{2}^{opt}$ and $\phi ^{opt}$, we can achieve the optimal
sensitivity of $\text{sg}$-detection that is
\begin{equation}
	\delta \phi _{\text{sg}}^{opt}=\frac{1+\sqrt{1-l}}{2\sqrt{(1-l)N}},
	\label{eq:6}
\end{equation}%
which depends on the loss rate $l$ and photon number $N$. \
\bigskip

\textbf{II. Difference-intensity (\text{df}) detection}

Difference-intensity detection measures the intensity difference of two
outputs ${a}_{3}$ and ${b}_{3}$, $\hat{N}_{ab}=\hat{a}_{3}^{\dag }\hat{a}%
_{3}-\hat{b}_{3}^{\dag }\hat{b}_{3}$. The phase sensitivity $\delta \phi
_{\text{df}}$ is obtained by (details see Appendix \ref{B2}.2)
\begin{equation}
	\begin{aligned}
		{\delta \phi _{\text{df}} =\frac{\left\langle \Delta \hat{N}_{ab}\right\rangle }{\left\vert \partial \left\langle \hat{N}_{ab}\right\rangle /\partial \phi\right\vert }=\frac{\sqrt{BN}}{2CN\left\vert \sin\phi \right\vert}}.
	\end{aligned}
\end{equation}
Different from the $\text{sg}$-detection, the optimal conditions for the best
sensitivity of $\text{df}$-detection $\delta \phi _{\text{df}}^{opt}$ are $R_{2}^{opt}=0.5$%
, $\phi ^{opt}$=$\pi /2$. Furthermore, the sensitivity $\delta \phi _{\text{df}}$
with $R_{2}^{opt}$ and $\phi^{opt}$ is given as
\begin{equation}
	\delta \phi _{\text{df}}=\sqrt{\frac{1-l+R_{1}l}{4(1-R_{1})R_{1}(1-l)}}\cdot \frac{1}{\sqrt{N}}.
	\label{eq:8}
\end{equation}%
Therefore, the optimal conditions corresponding to the minimum phase
sensitivity are
\begin{equation}
	\begin{aligned}
		&R_{1}^{opt}=\frac{\sqrt{1-l}}{1+\sqrt{1-l}}, R_{2}^{opt}=0.5,&\\
		&\phi^{opt}=\frac{\pi }{2},\frac{3\pi }{2}.& 
	\end{aligned}
\end{equation}%
Final optimal sensitivity of $\text{df}$-detection is
\begin{equation}
	\delta \phi _{\text{df}}^{opt}=\frac{1+\sqrt{1-l}}{2\sqrt{(1-l)N}}.
	\label{eq:10}
\end{equation}%
$R_{1}^{opt}$ and $\delta \phi _{\text{df}}^{opt}$ are the same as those of $\text{sg}$%
-detection. 

\bigskip

\textbf{III. Standard interferometric limit (SIL)}

Quantum Fisher information (QFI) of a lossy MZI has been discussed in Ref.\cite{29}.  The QFI results provide the optimal reflectivity $R_{1}$ for the first beam splitter, which is not relevant to the $R_{2}$ for the second beam splitter and the specific detection method. Therefore, there is an optimal phase estimation that can be achieved with the existing loss conditions (losses in two arms:$l_{a},l_{b}$), called standard interferometric limit (SIL). The phase sensitivity of SIL is given by
\begin{equation}
	\begin{aligned}
		\delta \phi _{\text{SIL}}=\frac{\sqrt{1-l_{a}}+\sqrt{1-l_{b}}}{2\sqrt{(1-l_{a})(1-l_{b})N}}
	\end{aligned}
	\label{eq:11}
\end{equation}%   
with the optimal  $R_{1}$
\begin{equation}
	\begin{aligned}
		R_{1}^{opt}=\frac{\sqrt{1-l_{a}}}{\sqrt{1-l_{a}}+\sqrt{1-l_{b}}}.
	\end{aligned}
\end{equation}%

$R_{1}^{opt}$ and $\delta \phi _{\text{SIL}}$ are the same as optimum cases of $\text{sg}$- and $\text{df}$-detections by setting $l_{a}=l,l_{b}=0$. 
From Eqs. (\ref{eq:6},\ref{eq:10},\ref{eq:11}), it is clear that SIL can be achieved with correct optimization on both detection methods.
\bigskip

\textbf{V. Analysis}

The theoretical comparison of the two detection methods before and after optimization is analyzed in detail in Fig. \ref{fig.2}.
Fig. \ref{fig.2}(a) gives the phase sensitivity with phase differences from 0 to 2$\pi$ when the loss rate is 0.9.
The minimum value of $\delta \phi _{\text{df}}$ is smaller than that of $\delta \phi _{\text{sg}}$ with balanced beam splitting ratios, showing the advantage of $\text{df}$-detection in lossy CMZI.
After optimizing the beam-splitting ratios, the sensitivity of two methods can reach SIL at their respective optimal phase differences.
As shown in Fig. \ref{fig.2}(b), the best phase sensitivity of CMZI with $\text{df}$-detection (red dashed line) is always better
than that of $\text{sg}$-detection (green dotted line) at all loss rates. Therefore,
the $\text{df}$-detection is more loss-tolerant than the $\text{sg}$-detection. More
importantly, the best phase difference $\phi ^{opt}$ of $\text{sg}$-detection
is different before and after optimization, while $\phi ^{opt}$ of $\text{df}$-detection is always
located at $\pi /2$. This means the optimization of MZI with the $\text{df}$
-detection is easier to operate by using phase-locking at $\pi /2$ in
practical application.

\begin{figure}
	\centering
	\includegraphics[width=0.8\columnwidth]{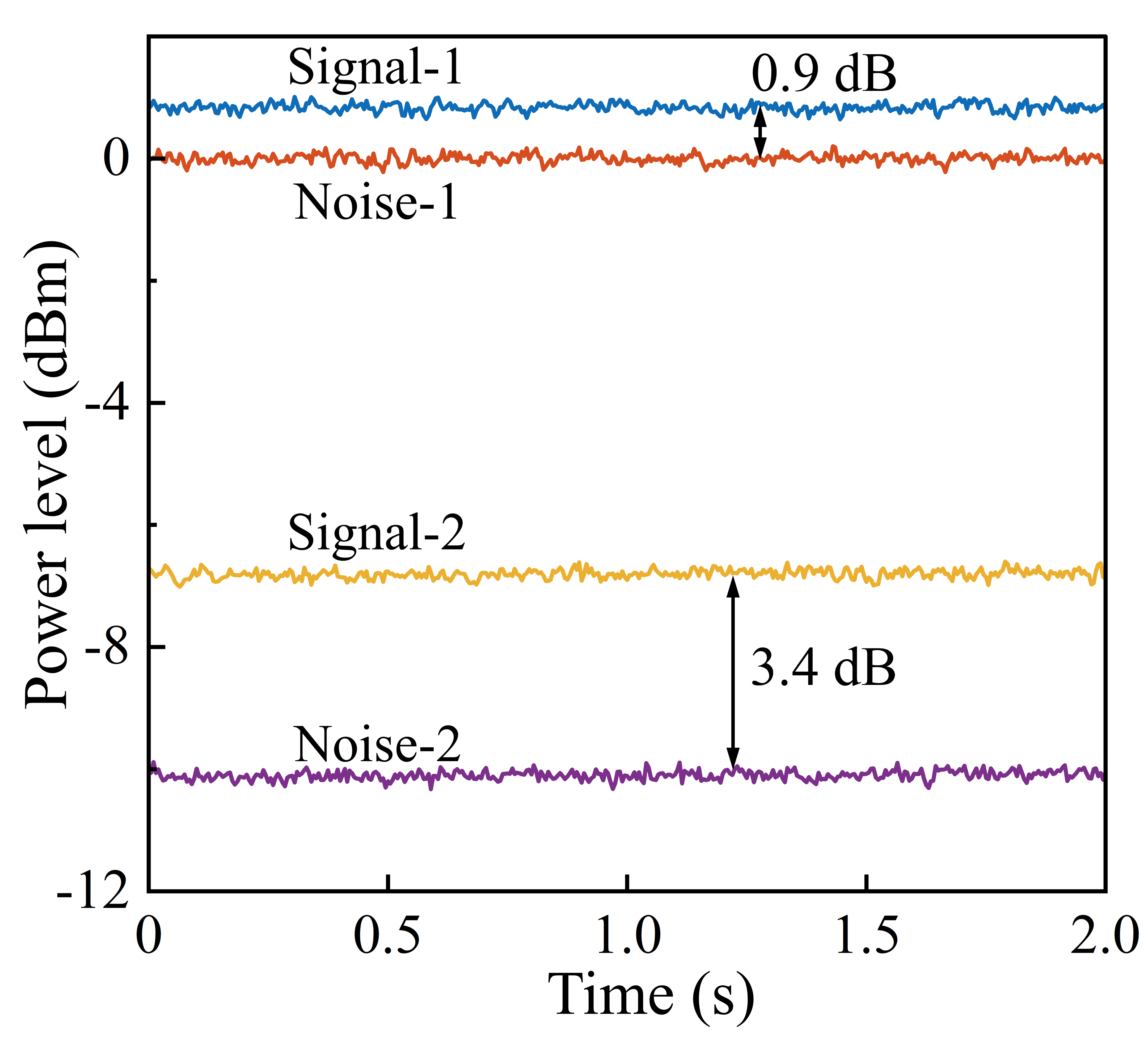}
	\caption{Signal and noise are measured using the $\text{df}$-detection at 1.5 MHz
		with a spectrum analyzer (SA). 1 and 2 represent the results of MZI with balanced reflectivities ($R_{1}=R_{2}=0.5$) and
		optimal reflectivities ($R_{1}=0.04,R_{2}=0.5$), respectively. A 2.5 dB
		signal-to-noise improvement is achieved via $R_{1}$ optimization. SA parameters: 1.5 MHz zero span with a
		resolution bandwidth (RBW) of 30 kHz and a video bandwidth (VBW) of 30 Hz.  Phase difference
		$\phi$ is locked at $\pi /2$. Loss rate is set to $l=0.998$.  Noise-1 is set to 0 dBm as a reference.}
	\label{fig.3}
\end{figure}

In this paper, we focus on sensitivity improvement via reflectivity
optimization. The sensitivity with both $\text{sg}$- and $\text{df}$-detections can be
improved by utilizing the optimal parameters ($R_{1}^{opt},R_{2}^{opt},\phi
^{opt}$) at all loss rates. The optimal sensitivities with two detection
schemes, $\delta \phi _{\text{df}}^{opt}$ and $\delta \phi _{\text{sg}}^{opt}$, are exactly the same (solid line in Fig. \ref{fig.2}(b)), and are also exactly overlapped with $\delta \phi _{\text{SIL}}$. The sensitivity improvement as a function of loss rate $l$ is given in Fig. \ref{fig.2}(c) normalized by CMZI with $\text{sg}$-detection ($\delta \phi _{\text{sg}}$) as a reference in 0 dBm. 

The sensitivity improvement increases with the loss. When the loss rate $l$ is close to 1, the sensitivity of CMZI with $\text{df}$-detection is better than that of $\text{sg}$-detection by 3 dB. Furthermore, the optimization of the reflectivities can bring a nearly 6 dB improvement with $\text{sg}$-detection. We choose $\text{df}$-detection in our experiments because the phase difference $\phi ^{opt}$ and reflectivity $R_{2}$ only need to be locked at $\pi/2$ and $0.5$ at any loss rate, and the optimal phase sensitivity is consistent with $\delta \phi _{\text{SIL}}$. When the losses change, we simply adjust $R_{1}$ to the optimum value, which facilitates practical applications.

\section*{Experiment}

\begin{figure*}
	\centering
	\includegraphics[width=2\columnwidth]{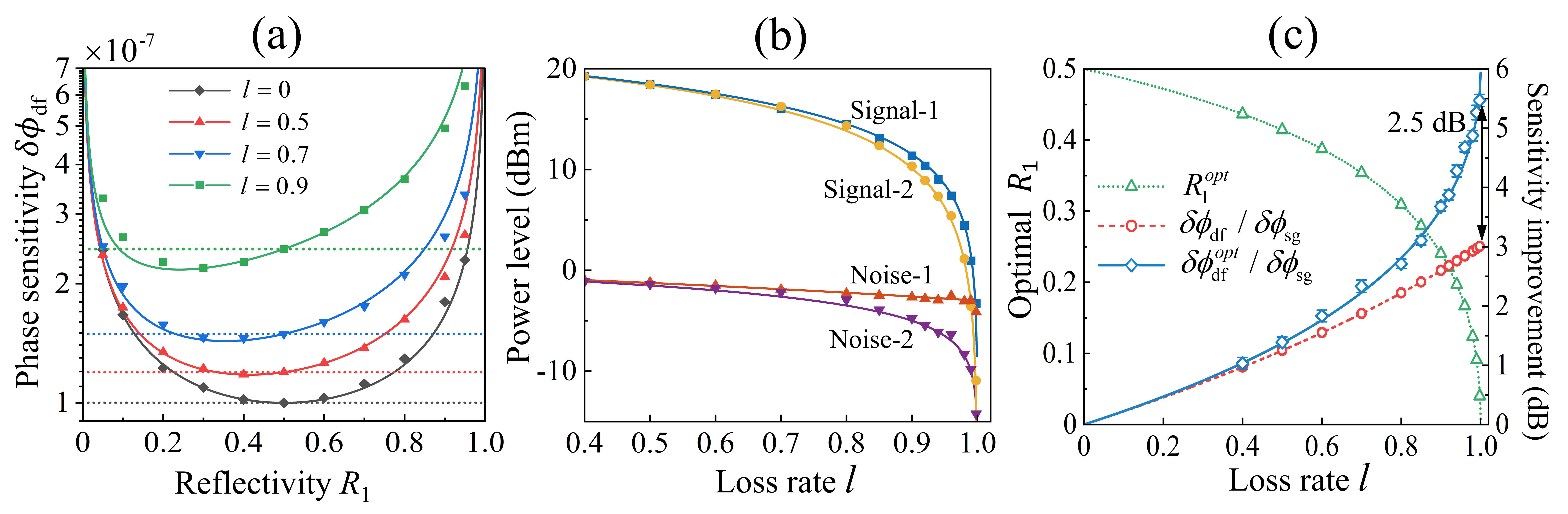}
	\caption{\textbf{(a)} Experimental data of phase sensitivity $\protect\delta
		\protect\phi $ versus $R_{1}$ at different loss rate $l$. Solid
		lines: theoretical prediction. Dotted lines: phase sensitivity of MZI with $R_{1}=R_{2}=0.5$. \textbf{(b)} Signal and
		noise measured with different loss rates $l$ when \textbf{(1)} $R_{1}=0.5$
		or \textbf{(2)} $R_{1}=R_{1}^{opt}$. \textbf{(c)} Optimal $R_{1}$ (left
		axis: dotted line and triangles) and sensitivity improvement
		(right axis: solid and dashed lines, diamonds and circles) as functions of the loss rate $l$. Lines: theoretical values; Scatters: experimental data.
		Sensitivity improvements are given with respect to $\delta\phi _{\text{sg}}$.
		$\delta\phi _{\text{df}}^{opt}$\ and $\delta\phi _{\text{df}}$\ are the 
		sensitivities of $\text{df}$-detection measured with optimal $R_{1}$ and non-optimal $R_{1}$=0.5,
		respectively.}
	\label{fig.4}
\end{figure*}
In the experiment, coherent light from a semiconductor laser as
the optical input field $a_{0}$ of the MZI. BS$_{1,2}$ are constructed as
shown in Fig. \ref{fig.1}(b,c). The input fields $a_{0}$ and $b_{0}$ with respective
horizontal and vertical polarizations are combined by a polarization beam
splitter (PBS$_1$) and then divided into two arms after passing through the half-wavelength plate (HWP$_1$) and PBS$_2$. The reflectivity $R_{1}$ is continuously adjusted by rotating HWP$_1$. A smaller $R_{1}$ value means more coherent light passing the loss path. The same design is applied for the BS$_2$. Considering the optimal conditions of the $\text{df}$-detection, HWP$_1$ is rotated to a specific degree and HWP$_2$ is fixed at 22.5$^{\circ}$ to achieve an adjustable $R_{1}$ and a constant $R_{2}=0.5$. An adjustable attenuator is placed in the signal arm $a$ of the interferometer for continuous simulation of the internal loss rate $l$. Since the output electrical signal of the $\text{df}$-detection is oddly symmetrical at phase difference $\phi=\pi/2$, part of it is used as an error signal for phase-locking (not shown in Fig. \ref{fig.1}(a)). The phase difference $\phi $ between two arms is always locked at optimal value $\pi /2$ during the measurement process. 
A small phase shift $\Delta \phi $ is introduced by a 1.5 MHz-frequency modulation signal using a piezoelectric ceramic mirror. Signal and noise are measured at 1.5 MHz using $\text{df}$-detection
by spectrum analyzer when the phase shift is modulated and unmodulated, respectively. As shown in Fig. \ref{fig.3}, the signal-1 (blue) and noise-1 (red)
curves represent the signal and noise of MZI with balanced $R_{1}$ when $l=0.998$. After optimizing $R_{1}$ to 0.04, the signal-to-noise ratio increased from 0.9 dB to 3.4 dB, although both signal and noise decreased. Finally, the phase sensitivity is obtained from the signal-to-noise conversion (see Appendix.\ref{A}) with a 2.5 dB sensitivity improvement.
To reflect the sensitivity improvement due to reflectivity $R_{1}$ optimization, we continuously measure the sensitivity as a function of $R_{1}$ at different losses ($l=0,0.5,0.7,0.9$) shown in Fig. \ref{fig.4}(a), and the results are in excellent agreement with Eq. \ref{eq:8}. 

This clearly shows that the optimal $R_{1}$ (the $R_{1}$ value at best sensitivity) for lossless MZI is the conventional reflectivity 0.5. 
The best sensitivity degrades with the increase of the loss rate.
Obviously, the loss results in a negative impact on the phase sensitivity. The optimal $R_{1}$ is smaller when the loss rate $l$ is larger.
Smaller $R_{1}$ means more photons passing through the lossy
arm, which results in degradation of both signal and noise but benefits the
sensitivity improvement. This is due to the decay rates of signal and noise
being different at different $R_{1}$. The optimal $R_{1}$ is the
balance between signal and noise decay. Furthermore, the decay rates of
signal and noise also change with the loss rate, resulting in the optimal $R_{1}
$ varying with $l$. This point can be seen from Eqs.(\ref{eq:6},\ref{eq:10}) and is also
shown in Fig. \ref{fig.4}(c). The optimal $R_{1}$ corresponding to the best sensitivity
decreases as the loss increases. To further explore the relationship between the sensitivity improvement and the loss rate, we measure the signal and noise with or without optimization of $R_{1}$ at loss rates from 0.4 to 0.998, as shown in Fig. \ref{fig.4}(b). The signal and noise both decrease as the loss increases, but the noise after optimization (Noise-2) decreases much faster, especially when the loss rate $l$ approaches 1. Thus, the sensitivity is improved more by optimizing $R_{1}$ at large losses.

In the experiment, we calculate the optimal $R_{1}$ values at different losses and adjust $R_{1}$ experimentally to measure the optimized phase sensitivity ($\delta \phi
_{\text{df}}^{opt}$). As shown in Fig. \ref{fig.4}(c), the $R_{1}^{opt}$ value decreases with
the increase of the loss rate. Compared with the sensitivity of unoptimized $\text{df}$-detection ($\delta\phi _{\text{df}}$), the optimal sensitivity ($\delta \phi _{\text{df}}^{opt}$) is improved at all loss rates and enhanced more at higher loss rate, benefited from the $R_{1}$ optimization. For example, $\delta\phi _{\text{df}^{opt}}$ achieves an optimization of 2.5 dB when l = 0.998, which equates to a 5.5 dB sensitivity improvement compared to the theoretical $\delta \phi_{\text{sg}}$. In general, the experimental results agree with the theoretical predictions well. 

\section*{Summary and Outlook}
In this paper, an MZI with variable beamsplitters is designed and demonstrated to improve the phase sensitivity with the existence of unbalanced losses. We theoretically give the optimal conditions and analyze the optimal phase sensitivity of two detections ($\text{sg}$-detection and $\text{df}$-detection) which both can saturate the optimal sensitivity$\sim$$\delta \phi _{\text{SIL}}$. When the loss rate $l$ is as high as 0.998, a sensitivity improvement of 2.5 dB is achieved in $\text{df}$-detection by optimizing $R_{1}$, which is equivalent to a 5.5 dB improvement in $\text{sg}$-detection. Clearly, in an MZI with balanced 50:50 beam splitters, it is difficult to improve sensitivity with large losses. Such optimization schemes should have a wide range of potential applications when unbalanced losses are unavoidable, e.g. optical hardware for networks, LISA schemes for the detection of gravitational waves, unbalanced MZI for continuous variable entanglement measurements, etc.

\section*{Acknowledgements}
This work is supported by the National Natural Science Foundation of China Grants No. 12274132, No. 11874152, No. 11974111, and No. 91536114;
Shanghai Municipal Science and Technology Major Project under Grant No. 2019SH-ZDZX01;
Innovation Program of Shanghai Municipal Education Commission No. 202101070008E00099; and Fundamental Research Funds for the Central Universities.

\textbf{Conflict of interest:} The authors declare that they have no conflict of interest.

\section*{Appendix}
\begin{appendix}
	
	\section{Phase sensitivity}	
	\label{A}
	\setcounter{equation}{0}%将公式编号归0
	\renewcommand{\theequation}{a\arabic{equation}}
	
	With a small phase shift $\Delta \phi $, the average of measurement operator $\hat{O}$ of MZI can be described as
	\begin{equation}
		\left\langle \hat{O}(\phi +\Delta \phi )\right\rangle \approx \left\langle
		\hat{O}(\phi )\right\rangle +\left\vert \frac{\partial \left\langle \hat{O}%
			\right\rangle }{\partial \phi }\right\vert \Delta \phi .
	\end{equation}%
	From the above equation, the signal and noise can be written as
	\begin{equation}
		\text{Signal}=\left\vert \frac{\partial \left\langle \hat{O}\right\rangle }{%
			\partial \phi }\right\vert \Delta \phi , \text{Noise}=\left\langle \Delta \hat{O}%
		\right\rangle ,
	\end{equation}%
	where $\left\langle \Delta \hat{O}\right\rangle =[\left\langle \hat{O}%
	^{2}\right\rangle -\left\langle \hat{O}\right\rangle ^{2}]^{1/2}$ is the
	standard deviation of $\hat{O}$. $\Delta \phi $ can only
	be detected when the signal is larger than the noise, therefore
	\begin{equation}
		\Delta \phi \geqslant \frac{\left\langle \Delta \hat{O}\right\rangle }{%
			\left\vert \partial \left\langle \hat{O}\right\rangle /\partial \phi
			\right\vert }.
	\end{equation}
	
	When we modulate the phase shift $\phi $ with a specific frequency signal, a
	small phase shift $\Delta \phi $ is introduced. Thus, the minimum detectable
	change in phase shift, called phase sensitivity $\delta \phi $, can be
	given as
	\begin{equation}
		\delta \phi =\Delta \phi _{\min }=\frac{\left\langle \Delta \hat{O}%
			\right\rangle }{\left\vert \partial \left\langle \hat{O}\right\rangle
			/\partial \phi \right\vert}=\frac{\text{Noise}}{\text{Signal}}\Delta \phi.
	\end{equation}
	
	\subsection{Losses in two arms ($l_{a}$, $l_{b}$)}
	\label{B}
	\setcounter{figure}{0}%将公式编号归0
	\renewcommand{\thefigure}{A\arabic{figure}}
	\begin{figure}
		\includegraphics[width=1\columnwidth]{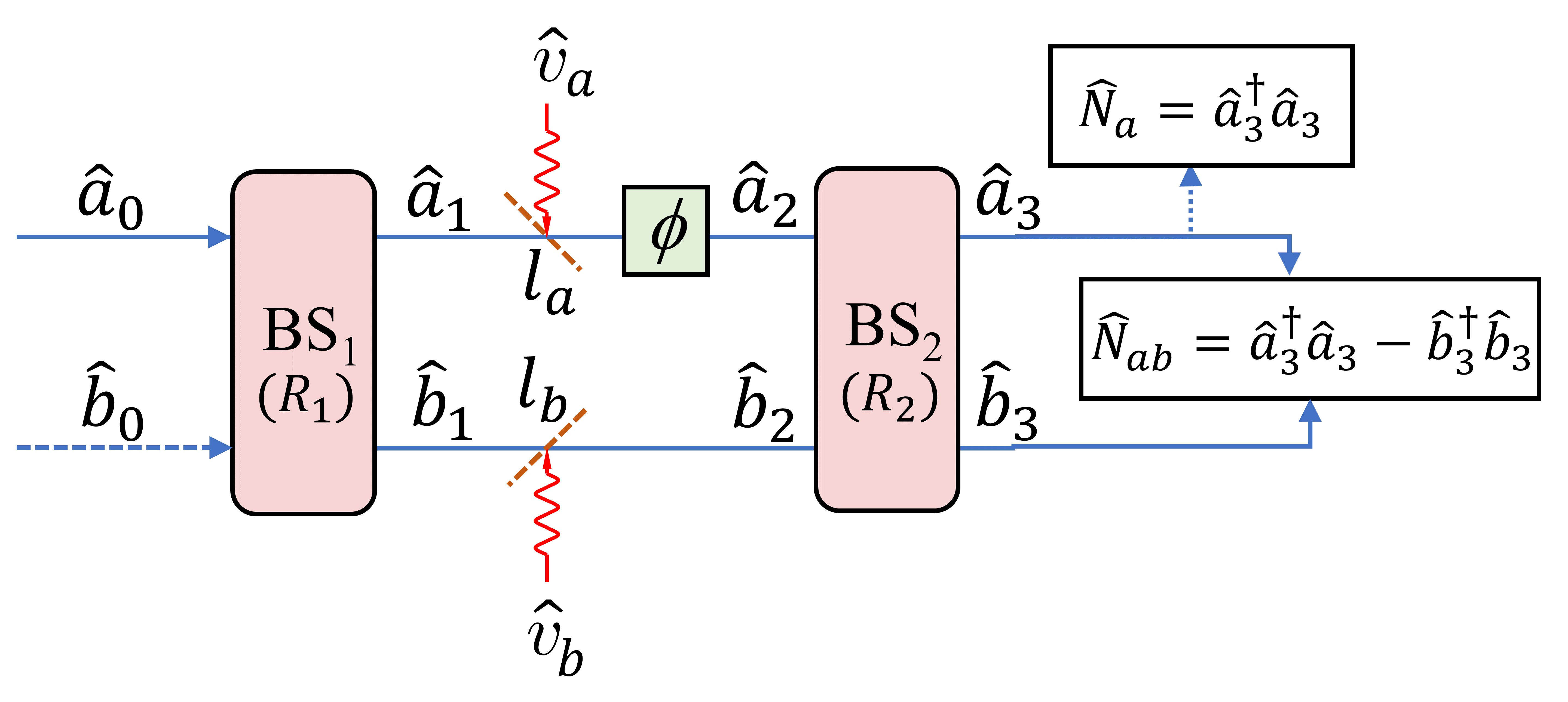}
		\caption{Model of internal losses on two arms of MZI. $l_{a}$,$l_{b}$:
			loss rates. $\hat{v}_{a},\hat{v}_{b}$: annihilation operators for vacuum
			fields introduced by losses $l_{a}$,$l_{b}$.}
		\label{fig.A1}
	\end{figure}
	
	An MZI model with double-arm internal losses ($l_{a}$,$l_{b}$) is illustrated in Fig. \ref{fig.A1}, which is more general than the simplified model in Fig. \ref{fig.1}(a) (only considering single-arm loss $l$ by setting $l_{a}=l, l_{b}=0$).
	The correspondence between the input and output states is as follows
	\begin{equation}
		\begin{aligned}
			\hat{a}_{3}& =\sqrt{1-R_{2}}\hat{a}_{2}+i\sqrt{R_{2}}\hat{b}_{2}&  
			\\
			&=k_{1}\hat{a}_{0}+k_{2}\hat{b}_{0}+k_{3}\hat{v}_{a}+k_{4}\hat{v}_{b},& \\
			\hat{b}_{3}& =\sqrt{1-R_{2}}\hat{b}_{2}+i\sqrt{R_{2}}\hat{a}_{2}&  
			\\
			&=h_{1}\hat{a}_{0}+h_{2}\hat{b}_{0}+h_{3}\hat{v}_{a}+h_{4}\hat{v}_{b},&
		\end{aligned}
	\end{equation}%
	where the assumed coefficients are
	\begin{equation}
		\begin{aligned}
			&k_{1} =\sqrt{(1-R_{1})(1-R_{2})(1-l_{a})}e^{i\phi }-\sqrt{R_{1}R_{2}(1-l_{b})},& \\
			&k_{2} =i\sqrt{R_{1}(1-R_{2})(1-l_{a})}e^{i\phi }+i\sqrt{(1-R_{1})R_{2}(1-l_{b})},& \\
			&k_{3} =i\sqrt{(1-R_{2})l_{a}}e^{i\phi},& \\
			&k_{4} =-\sqrt{R_{2}l_{b}},& \\
			&h_{1} =i\sqrt{R_{1}(1-R_{2})(1-l_{b})}+i\sqrt{(1-R_{1})R_{2}(1-l_{a})}e^{i\phi },& \\
			&h_{2} =\sqrt{(1-R_{1})(1-R_{2})(1-l_{b})}-\sqrt{R_{1}R_{2}(1-l_{a})}e^{i\phi },& \\
			&h_{3} =-\sqrt{R_{2}l_{a}}e^{i\phi},&\\
			&h_{4} =i\sqrt{(1-R_{2})l_{b}}.& 
		\end{aligned}
	\end{equation}
	
	\subsubsection*{\begin{center}
			1. Single-intensity detection
			\label{B1}
	\end{center}}
	
	When intensity detection is performed at a detector, the signal ($\left\vert \partial \left\langle \hat{N}_{a}\right\rangle /\partial \phi
	\right\vert\Delta \phi$), noise ($\left\langle \Delta \hat{N}_{a}\right\rangle$), and
	phase sensitivity ($\delta \phi _{\text{sg}}$) can be obtained by the following
	calculation
	\begin{equation}
		\begin{aligned}
			&\left\langle \Delta \hat{N}_{a}\right\rangle =[\left\langle \hat{N}{a}^{2}\right\rangle -\left\langle \hat{N}_{a}\right\rangle ^{2}]^{1/2}=\sqrt{(B-C\cos \phi)N},&  \\
			&\left\vert \partial \left\langle \hat{N}_{a}\right\rangle /\partial \phi\right\vert =CN\left\vert \sin \phi \right\vert,& \\
			&\delta \phi _{\text{sg}}=\frac{\left\langle \Delta \hat{N}_{a}\right\rangle }{\left\vert \partial \left\langle \hat{N}_{a}\right\rangle /\partial \phi\right\vert }=\frac{\sqrt{B-C\cos \phi}}{C\left\vert \sin \phi \right\vert}\cdot \frac{1}{\sqrt{N}},&
		\end{aligned}
		\label{eq:a7}
	\end{equation}
	
	with
	\begin{equation}
		\begin{aligned}
			&\left\langle \hat{N}_{a}\right\rangle =\left\langle \hat{a}_{3}^{\dag }\hat{a}_{3}\right\rangle =k_{1}^{\ast }k_{1}N,&   \\
			&\left\langle \hat{N}_{a}^{2}\right\rangle =\left\langle \hat{a}_{3}^{\dag }\hat{a}_{3}\hat{a}_{3}^{\dag }\hat{a}_{3}\right\rangle =(k_{1}^{\ast}k_{1})^{2}N^{2}+k_{1}^{\ast }k_{1}N,&   \\
			&B=(1-R_{1})(1-R_{2})(1-l_{a})+R_{1}R_{2}(1-l_{b}),&   \\
			&C=2\sqrt{R_{1}R_{2}(1-R_{1})(1-R_{2})(1-l_{a})(1-l_{b})}.&  
		\end{aligned}
	\end{equation}
	The Phase sensitivity $\delta \phi _{\text{sg}}$ is a function of the parameters $%
	R_{1},R_{2},l_{a},l_{b}$ and $\phi$. From Eq. (\ref{eq:a7}), we can determine the
	optimization condition for the minimum phase sensitivity as
	\begin{equation}
		\begin{aligned}
			&R_{1}^{opt}=R_{2}^{opt}=\frac{\sqrt{1-l_{a}}}{\sqrt{1-l_{a}}+\sqrt{1-l_{b}}},&\\
			&\phi ^{opt}=\arccos (\frac{B-\sqrt{B^{2}-C^{2}}}{C})=0,2\pi.& 
		\end{aligned}
		\label{eq:a9}
	\end{equation}
	
	\subsubsection*{\begin{center}
			2.	Difference-intensity detection
			\label{B2}
	\end{center}}
	
	When using two detectors for difference-intensity detection, the
	signal, noise and phase sensitivity $\delta \phi _{\text{df}}$ can be expressed as
	
	\begin{equation}
		\begin{aligned}
			&\left\langle \Delta \hat{N}_{ab}\right\rangle =[\left\langle \hat{N}_{ab}^{2}\right\rangle -\left\langle \hat{N}_{ab}\right\rangle ^{2}]^{1/2}=\sqrt{BN},& \\
			&\left\vert \partial \left\langle \hat{N}_{ab}\right\rangle /\partial \phi\right\vert=2CN\left\vert \sin \phi \right\vert ,& \\
			&\delta \phi _{\text{df}} =\frac{\left\langle \Delta \hat{N}_{ab}\right\rangle }{\left\vert \partial \left\langle \hat{N}_{ab}\right\rangle /\partial \phi\right\vert }=\frac{\sqrt{B}}{2C\left\vert \sin\phi \right\vert}\cdot \frac{1}{\sqrt{N}}.&
		\end{aligned}
	\end{equation}
	with
	\begin{equation}
		\begin{aligned}
			\left\langle \hat{N}_{ab}\right\rangle &=\left\langle \hat{a}_{3}^{\dag }\hat{a}_{3}-\hat{b}_{3}^{\dag }\hat{b}_{3}\right\rangle =\left\langle A_{11}\hat{a}_{0}^{\dag }\hat{a}_{0}\right\rangle,& \\
			\left\langle \hat{N}_{ab}^{2}\right\rangle &=A_{11}^{2}\left\langle \hat{a}_{0}^{\dag }\hat{a}_{0}\hat{a}_{0}^{\dag }\hat{a}_{0}\right\rangle+A_{12}A_{21}\left\langle \hat{a}_{0}^{\dag }\hat{b}_{0}\hat{b}_{0}^{\dag }\hat{a}_{0}\right\rangle& \\
			&+A_{13}A_{31}\left\langle \hat{a}_{0}^{\dag }\hat{v}_{a}\hat{v}_{a}^{\dag} \hat{a}_{0}\right\rangle +A_{14}A_{41}\left\langle \hat{a}_{0}^{\dag }\hat{v} _{b}\hat{v}_{b}^{\dag }\hat{a}_{0}\right\rangle& \\
			&=A_{11}^{2}N^{2}+(A_{11}^{2}+A_{12}A_{21}+A_{13}A_{31}+A_{14}A_{41})N,& \\
			A_{ij} &=k_{i}^{\ast }k_{j}-h_{i}^{\ast }h_{j}(i,j=1,2,3,4).&
		\end{aligned}
	\end{equation}
	
	Unlike $\text{sg}$-detection, the phase sensitivity of difference-intensity
	detection ($\delta \phi _{\text{df}}$) always takes the minimum value at $\phi =%
	\frac{\pi }{2}$ and $R_{2}=0.5$ with reference to Eq. (\ref{eq:a12}).
	\begin{equation}
		\begin{aligned}
			\frac{\partial (\delta \phi _{\text{df}})}{\partial \phi }\equiv0|_{\phi =\frac{\pi }{2}},\frac{\partial (\delta \phi _{\text{df}})}{\partial R_{2}}\equiv0|_{R_{2}=0.5}.
		\end{aligned}
		\label{eq:a12}
	\end{equation}
	After setting phase shift $\phi $ to $\frac{\pi }{2}$ and reflectivity $%
	R_{2} $ to $0.5$, the simplified phase sensitivity is
	\begin{equation}
		\delta \phi _{\text{df}}=\sqrt{\frac{1-l_{a}+R_{1}(l_{a}-l_{b})}{%
				4(1-R_{1})R_{1}(1-l_{a})(1-l_{b})}}\cdot \frac{1}{\sqrt{N}}.
		\label{eq:a13}
	\end{equation}%
	Using Eq. (\ref{eq:a13}), we can find the conditions for optimal phase sensitivity
	\begin{equation}
		\begin{aligned}
			&R_{1}^{opt}=\frac{\sqrt{1-l_{a}}}{\sqrt{1-l_{a}}+\sqrt{1-l_{b}}}, R_{2}^{opt}=0.5,&\\
			&\phi^{opt}=\frac{\pi }{2},\frac{3\pi}{2}.&
		\end{aligned}
		\label{eq:a14}
	\end{equation}
	When both detection methods are under optimal conditions (Eq. \ref{eq:a9}, Eq. \ref{eq:a14}), their best phase sensitivities reach SIL.
	\begin{equation}
		\begin{aligned}
			\delta \phi _{\text{df}}^{opt}=\delta \phi _{\text{sg}}^{opt}=\delta \phi _{\text{SIL}}=\frac{\sqrt{1-l_{a}}+\sqrt{1-l_{b}}}{2\sqrt{(1-l_{a})(1-l_{b})N}}.
		\end{aligned}
		\label{eq:a15}
	\end{equation}% 
	The unbalanced loss has the same characteristics for the optimization of sensitivity, whether on a single arm or double arms. For convenience, we consider the case of loss just on a single arm in this paper ($l_{a}=l,l_{b}=0$). 
	
	\subsection{Measurement}
	The current of the $\text{df}$-detection is divided into two ways, 
	one as an error signal to lock the phase difference of the MZI at $\pi/2$, and the other for signal measurement.
	\label{C}
	\begin{figure}
		\includegraphics[width=1\columnwidth]{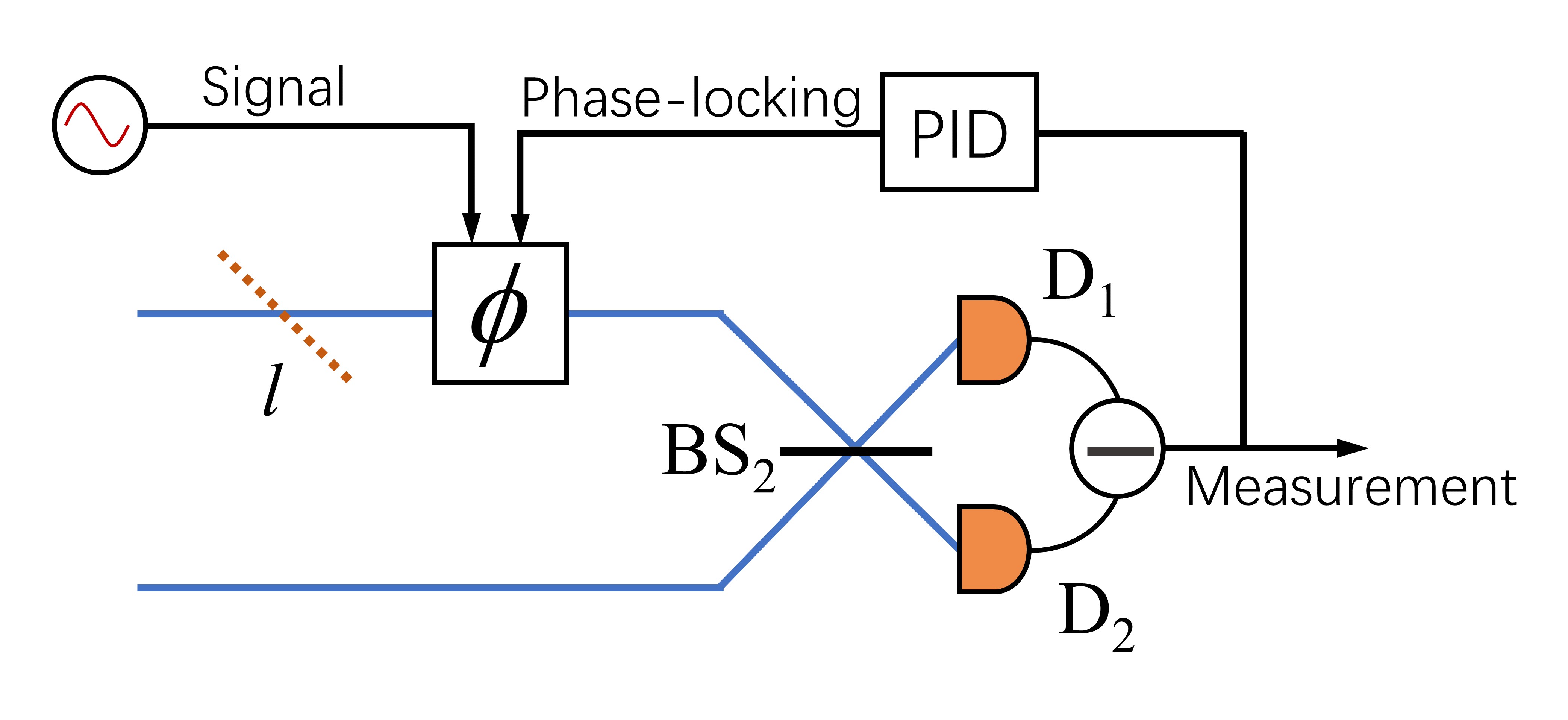}
		\caption{Experimental details. $D_1$, $D_2$: detectors.}
		\label{fig.A2}
	\end{figure}
	
\end{appendix}

\end{document}